\documentstyle[preprint,aps,epsfig]{revtex}

\def\be{\begin{equation}}
\def\ee{\end{equation}}
\def\bq{\begin{eqnarray}}
\def\eq{\end{eqnarray}}
\def\ep{\epsilon}

\begin{document}
\draft

\title{Many-body dipole-induced dipole model \\ for electrorheological fluids}
\author{J. P. Huang$^{1,2}$, K. W. Yu$^{1}$}
\address{$^1$Department of Physics, The Chinese University of Hong Kong,
 Shatin, NT, Hong Kong \\
 $^2$Laboratory of Computational Engineering, Helsinki University of Technology, 
 P. O. Box 9203, FIN-02015 HUT, Finland}
\maketitle

\begin{abstract}
Theoretical investigations on electrorheological (ER) fluids usually 
rely on computer simulations. 
An initial approach for these studies would be the point-dipole (PD) 
approximation, which is known to err considerably when the particles 
approach and finally touch due to many-body and multipolar interactions.
Thus various work attempted to go beyond the PD model.
Being beyond the PD model, previous attempts have been restricted to either 
local-field effects only or multipolar effects only, but not both.
For instance, we recently proposed a dipole-induced-dipole (DID) model 
which is shown to be both more accurate than the PD model and easy to use.
This work is necessary because the many-body (local-field) effect is included
to put forth the many-body DID model.
The results show that the multipolar interactions can indeed be dominant 
over the dipole interaction, while the local-field effect may yield 
an important correction.
\end{abstract}
\vskip 5mm 

\section{Introduction}

An electrorheological (ER) fluid consists of polarizable particles in a nonconducting host fluid. If an external electric field is applied to an ER fluid, the particles aggregate and form chains parallel to the applied field. These chains may further aggregate to columnlike structures \cite{Halsey}. To discuss ER effect, early theory is often based on the point-dipole (PD) approximation, e.g. \cite{Tao91}. Also, the PD approximation is often adopted in computer simulation \cite{Klingen} because it is simple and easy to use. Since many-body and multipolar interactions between particles have been neglected, the predicted strength of ER effect is of an order lower than the experimental results. Hence, much work has been done to sort out more accurate models \cite{K91,Davis92,Cl93,Yu2000}. Although those methods are accurate, they are relatively complicated to use in dynamic simulation of ER fluids \cite{Siu01}.

In a recent paper, we put forth the dipole-induced dipole (DID) model to improve
the point-dipole (PD) model \cite{Yu00}. In that work, we considered pair interaction between
polarized dielectric particles through the multiple image formula, 
but neglected the many-body interactions because 
it was believed that the multipolar interactions can be dominated over the many-body 
(local-field) effects. The claim was 
based on the calculation for a pair of particles. We should really question the 
quantitative accuracy of this claim, since in ER fluids chain and sheet 
structures are known to occur. In a long particle chain the local field 
at a dipole site is much larger than for a pair of dipoles, whereas for 
multipolar interactions this is probably not the case. In fact, this 
issue has already been quantified, for particles with large dielectric 
contrast with the liquid phase, by Martin and Anderson \cite{Martin-3}.
Thus we have to do some additional analysis to validate the claim. 

As an initial model, we may adopt an effective dipole factor in the 
multiple image formula \cite{Yu00}. More precisely, we regard that each polarized 
particle in an ER suspension is embedded in an effective medium with an 
effective dielectric constant $\ep_e$. Thus, the usual dipole factor 
$b = (\ep_1 - \ep_2)/(\ep_1 + 2\ep_2)$ in the multiple image theory should be 
replaced by $b' = (\ep_1 - \ep_e)/(\ep_1 + 2\ep_e)$ \cite{PLA02}, where $\ep_e$ is the 
effective dielectric constant of the surrounding medium, which is 
conveniently calculated in the Maxwell-Garnett approximation (MGA). 
Here $\ep_1$ and $\ep_2$ are the dielectric constants of the suspended particles 
and host fluid, respectively.
However, the situation is further complicated by the fact that the 
suspended particles in an ER fluid usually form anisotropic structures, 
e.g. chain and sheet structures, along the applied field and that $\ep_e$ 
can be anisotropic. Thus, $\ep_e$ should be calculated within the anisotropic 
MGA \cite{Lo,PRE01}. A preliminary calculation shows that $\ep_e$ should be increased 
and $b'$ should be decreased as compared to $b$. Thus the DID interaction would 
be reduced. But of course the actual reduction should depend on the 
local-field factor of the anisotropic structures. Thus the many-body 
effects have been included through the volume-fraction dependent $\ep_e$.

So, in the presence of the surrounding particles, the interaction between 
two particles will be modified. By means of an effective dielectric 
constant, we can derive an analytic formula for the many-particle DID 
model. And more importantly, we can access the importance of the 
many-body (local-field) effects against the multipolar interactions. 
We believe this work is necessary because being beyond the fixed dipole 
model, the previous work dealed with either local-field effects only 
\cite{Tao96} or multipolar effects only, but not both.

\section{Formalism}

We concentrate on the case where highly polarized dielectric particles of diameter $d$,
dielectric constant $\ep_1$ are embedded in a host fluid of $\ep_2$. 

In the dilute limit, the dipole factor for an isolated particle may be given by 
\be
b=\frac{\ep_1-\ep_2}{\ep_1+2\ep_2}.
\label{b0}
\ee
On the other hand, based on the multiple image method \cite{Yu00} the dipole factor for a pair of particles with separation $s$ has the following form:
\begin{eqnarray}
b_T ^*&=&b\sum_{n=0}^{\infty}(-b)^n[\frac{\sinh \alpha}{\sinh (n+1)\alpha}]^3,\nonumber\\
b_L ^*&=&b\sum_{n=0}^{\infty}(2b)^n[\frac{\sinh \alpha}{\sinh (n+1)\alpha}]^3,
\end{eqnarray}
for transverse field (T) and longitudinal field (L) case, respectively, where $\alpha$ satisfies the relation $\cosh \alpha=s/d$.

When a suspension containing many dielectric particles is subjected
to an intense electric field, the induced dipole moments may cause the 
particles to form chains along the applied field, resulting in complex
anisotropic structures. In this case, we may
invoke the Maxwell-Garnett approximation (MGA) for anisotropic 
composites \cite{Lo,PRE01} to obtain the effective dielectric function $\ep_e$ of the system.
For the transverse field case when the electric field is applied perpendicular to
the uniaxial anisotropic axis, the MGA has the form 
 \be
\frac{\ep_{eT}-\ep_2}{\beta_{T}\ep_{eT}+(3-\beta_{T})\ep_2}=
f\frac{\ep_1-\ep_2}{\ep_1+2\ep_2} ,
 \ee 
whereas for a longitudinal field case when the field is applied along the uniaxial anisotropic axis, the MGA reads
\begin{equation}
\frac{\ep_{eL}-\ep_2}{\beta_{L}\ep_{eL}+(3-\beta_L)\ep_2}=f\frac{\ep_1-\ep_2}{\ep_1+2\ep_2},
\end{equation}
where $\beta_{T}$ and $\beta_{L}$ denote the local field factors perpendicular
and parallel to the uniaxial anisotropic axis, and $f$ the volume fraction of particles.
These local field factors are defined as the ratio of the local field
in the particles to the Lorentz cavity field \cite{Lo}.
For isotropic composites, $\beta_L=\beta_T=1$, while both $\beta_L$ and 
$\beta_T$ will deviate from unity for an anisotropic distribution of
particles in composites.
These $\beta$ factors have been evaluated in a tetragonal lattice of
dipole moments \cite{Lo} and in various field-structured composites 
\cite{Martin-3} and they satisfy the sum rule $\beta_{L}+2\beta_{T}=3.$
In this case, the dipole factor for an individual particle may have this form \cite{PLA02}
\be
b^{'}=\frac{\ep_1-\ep_e}{\ep_1+2\ep_e},
\label{b'}
\ee
Note, to obtain this equation, we have replaced $\ep_2$ in Eq.(\ref{b0}) with $\ep_e$. That is, we assume the particle to be embedded in an effective medium.
Thus the many-body (local-field) effect is included already in Eq.(\ref{b'}). So far, to put forth the many-body DID
model, we may consider a pair of dielectric spheres as well. The spheres are placed in an effective medium. A constant electric field ${\bf E}_0=E_0\hat{{\bf z}}$ is applied to the spheres, which contribute to each sphere a dipole moment given by $p_{10}$ and $p_{20}(=p_{10}=\ep_eE_0d^3b^{'}/8)$. Thus the multiple image formula can be developed in a similar way \cite{Yu00}. The dipole moment $p_{10}$ induces an image dipole $p_{11}$ in sphere 2, while $p_{11}$ induces yet another image dipole in sphere 1. As a result, multiple images are formed. Similarly, $p_{20}$ induces an image $p_{21}$ inside sphere 1, and hence another infinite series of image dipoles are formed. Then we may obtain the sum of dipole moment inside each sphere, and hence the desired expressions for dipole factors:
\begin{eqnarray}
b'{} _T ^{*}&=&b^{'}\sum_{n=0}^{\infty}(-b^{'})^n[\frac{\sinh \alpha}{\sinh (n+1)\alpha}]^3\nonumber\\
b'{} _L ^{*}&=&b^{'}\sum_{n=0}^{\infty}(2b^{'})^n[\frac{\sinh \alpha}{\sinh (n+1)\alpha}]^3
\end{eqnarray}
for transverse field and longitudinal field case, respectively. The two equations
are nontrivial results indeed, which include the multiple image as well as many-body (local-field) effect.

To discuss the interparticle force, we may take one step forward to calculate the force between two dielectric spheres. We have already obtained the dipole factors $b^{'*}$ and $b^{*}$ and hence the dipole moments, and the force can be calculated by an energy approach. In this case, the dipole energy ($En$) of a pair of particles may be determined by the dot product of dipole moment and the electric field, and hence the force between a pair of particles is given by the derivative of dipole energy with respect to separation, namely $-{\rm d}En/{\rm d}s$.
So, based on this relation, we may obtain the expressions for interparticle forces, respectively:
\begin{eqnarray}
F_T^*&=&-(1/8)\ep_2E_0^2d^3b\sum_{n=0}^{\infty}(-b)^n\Phi,\nonumber\\
F_L^*&=&-(1/8)\ep_2E_0^2d^3b\sum_{n=0}^{\infty}(2b)^n\Phi,\nonumber\\
F'{}_T ^*&=&-(1/8)\ep_{eT}E_0^2d^3b_T' \sum_{n=0}^{\infty}(-b_T' )^n\Phi,\nonumber\\
F'{}_L ^*&=&-(1/8)\ep_{eL}E_0^2d^3b_L' \sum_{n=0}^{\infty}(2b_L' )^n\Phi,
\end{eqnarray}
with
\be
\Phi=\frac{\sinh \alpha \cosh \alpha \sinh[(n+1)\alpha]-(n+1)\sinh^2\alpha \cosh[(n+1)\alpha]}{(d/3)\sinh^4[(n+1)\alpha]},
\ee
where $F_T ^*$ ($F_L ^*$) indicates the interparticle force between a pair of particles in fluid host for transverse (longitudinal)  field case, and $F'{}_T ^*$ ($F'{}_L ^*$) in an effective medium for transverse (longitudinal)  field case. It is shown that the multipolar effect as well as many-body effects have been taken into account. Setting $n$ up to $1$ will yield directly the point-dipole forces, namely $F_T$ and $F_L$.

So far, we have put forth many-body DID model. We are now in a position to do some numerical calculations to discuss the $\beta$ effect on reduction factor $R=b^{'*}/b^*$, which indicates the correction of the many-body (local-field) effect on the multiple image effect, as well as on the ratio of $b^*$ or $b^{'*}$ to $b$. Then, to investigate such effect on interparticle force, we also investigate the interparticle force normalized by point-dipole force.

\section{Numerical Results}

For numerical calculations, we choose $\ep_1=10\ep_0$, $\ep_2=2.5\ep_0$, $d=2.0\times 10^{-7}$ and $s/d=1.01$, where $\ep_0$ is the dielectric constant of free space. From the numerical calculations, we find that a domination of the multipolar interactions is possibly over the many-body (local-field) effects. More precisely, in Fig.1, for the transverse field case, the correction due to the local-field effect is always small, because the reduction factor is close to unity. However, for the longitudinal field case, it is evident that the local-field effect can be large, especially at high volume fractions and/or large $\beta_L$. In which case, the correction due to local field cannot be neglected. In Fig.2, we compute the ratio of $b^*$ or $b^{'*}$ to $b$ versus the local field factor $\beta_L$ for different volume fractions.
In Fig. 3, we discuss the dependence of $\beta_L$ on the interparticle force normalized by point-dipole force for different volume fractions. From Figs.2 and 3, it is obvious that the multipolar interactions can indeed be dominant over the dipole interaction, while the local-field effect may yield an important correction.

The force expression is reasonable, as the $n=0$ term of Eq.(8) vanishes while
the $n=1$ term gives the correct dipole interaction force $r^{-4}$ dependence.
The divergence of the summation may be found for the longitudinal field case, 
due to the small distance $s/d=1.01$ being used. 

From the results, it is clear that the correction due 
to many-body (local-field) effects can be very large, especially at large 
volume fractions. Nevertheless, for small volume fractions, the 
correction due to the local field can be neglected.
The reduction of the magnitude of the effective interaction between two 
particles can be understood in a simple geometry in which the two particles 
interact in the presence of a third particle. An average over all possible 
positions of the third particle gives the desired reduction which is 
valid both for longitudinal and transverse fields. 
So, we believe our results are correct. 

\section{Discussion and conclusion}

Here a few comments are in order. The ER effect arises from the polarization forces produced by a mismatch in the dielectric constants of the disperse and continuous phases. Typical ER fluids contain a disperse phase with volume fractions in the range of $0.05\sim0.50$. In this sense, it is reasonable to set $f=0.1, 0.2, 0.3$ in our calculations. On the other hand, we set $\ep_2=2.5\ep_0$, which is actual value for silicon oil and commonly employed as the host fluid. Also, the disperse phases are generally composed of solid, nonconducting, or semiconducting materials with dielectric constants in the range $2\sim40$. Thus, the choice $\ep_1=10\ep_0$ is reasonable in our numerical calculations.
In addition, the longitudinal local field factor $\beta_L$ should satisfy $1\le\beta_L\le3$, in order to ensure that $\beta_T\ge 0$.
Regarding the separation between the pair of particles, various different values have been used as well, e.g. $s/d=1.03$ etc., and similar results are obtained (not shown here). However, for the case where the separation is too large, e.g. $s/d>2$, the multiple image effect will become small enough to be neglected \cite{PRE2}.

Our many-body DID model may be used in various studies of the behavior of ER fluids, e.g., in a theory describing the ground state of ER solids or in a computer simulation of ER fluids. By including the many-body effects, the DID model can be used with a higher accuracy. Lastly, it is straightforward to extend our model to deal with polydisperse ER fluids, and to discuss the magnetorheological suspensions.

In summary, we have developed a many-body DID model by including the many-body (local-field) effect, to compute the interparticle force for an ER fluid. We applied the formalism to a pair of touching spherical particles embedded in an effective medium, and calculated the force as a function of the longitudinal local-field factor. The result show that the multipolar interactions can indeed be dominant over the dipole interaction, while the local-field effect may yield an important correction, especially at larger volume fractions.

\section*{Acknowledgments}

This work was supported by the Research Grants Council of the Hong Kong SAR
Government under project number CUHK 4245/01P. 
J. P. H. thanks Dr. M. Karttunen for his invitation to the Laboratory of Computational Engineering, Helsinki University of Technology.

\begin{figure}[h]
\caption{The reduction factor versus local field factor $\beta_L$ for different volume fractions, $f=0.1$, $0.2$, and $0.3$.}
\end{figure}

\begin{figure}[h]
\caption{The ratio of $b^*$ or $b^{'*}$ to $b$ versus local field factor $\beta_L$ for different volume fractions, $f=0.1$, $0.2$, and $0.3$.}
\end{figure}

\begin{figure}[h]
\caption{The ratio of interparticle force to point-dipole force for different volume fractions, $f=0.1$, $0.2$, and $0.3$.}
\end{figure}

\newpage
\centerline{\epsfig{file=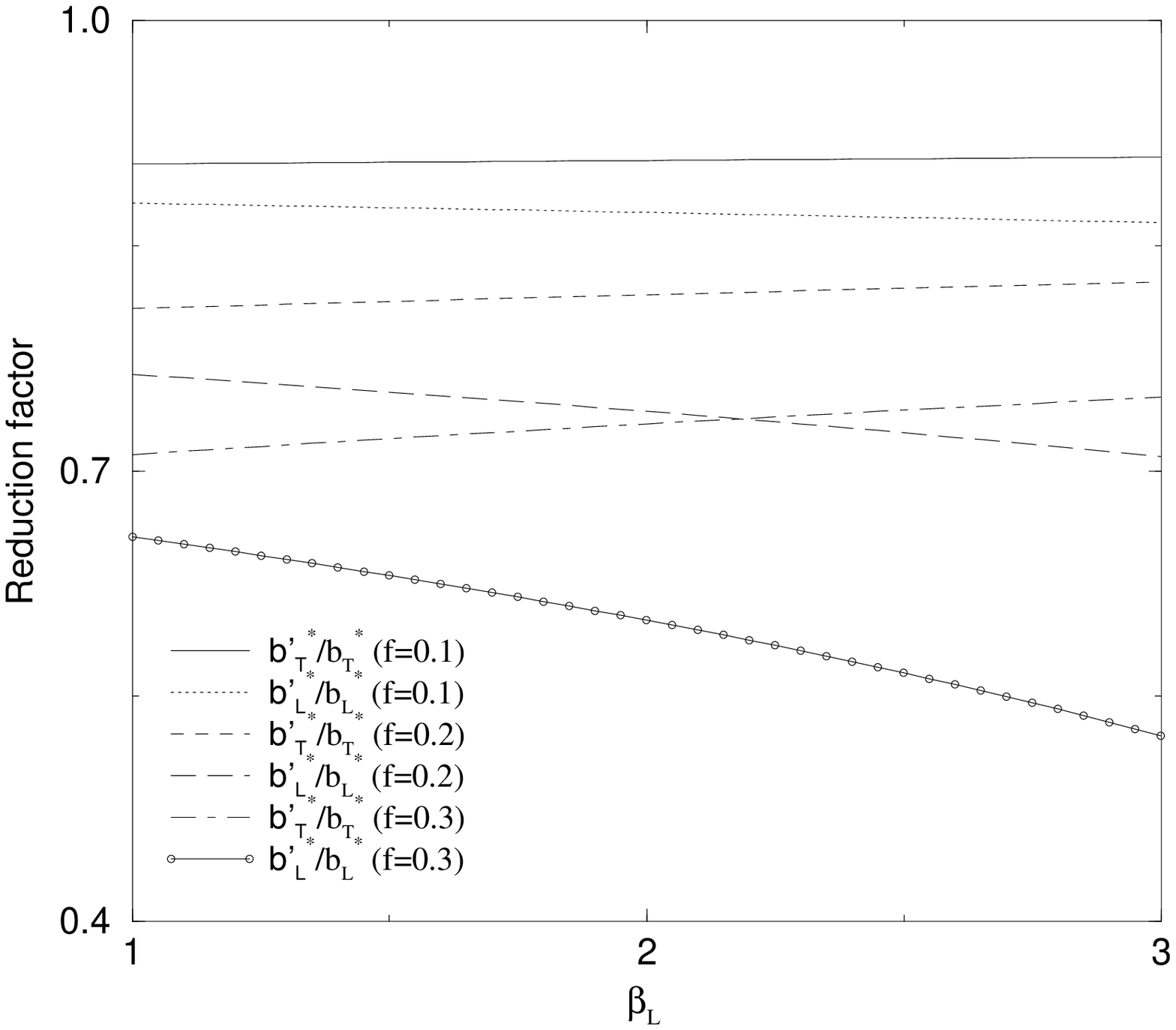,width=\linewidth}}
\centerline{Fig.1/Huang and Yu}

\newpage
\centerline{\epsfig{file=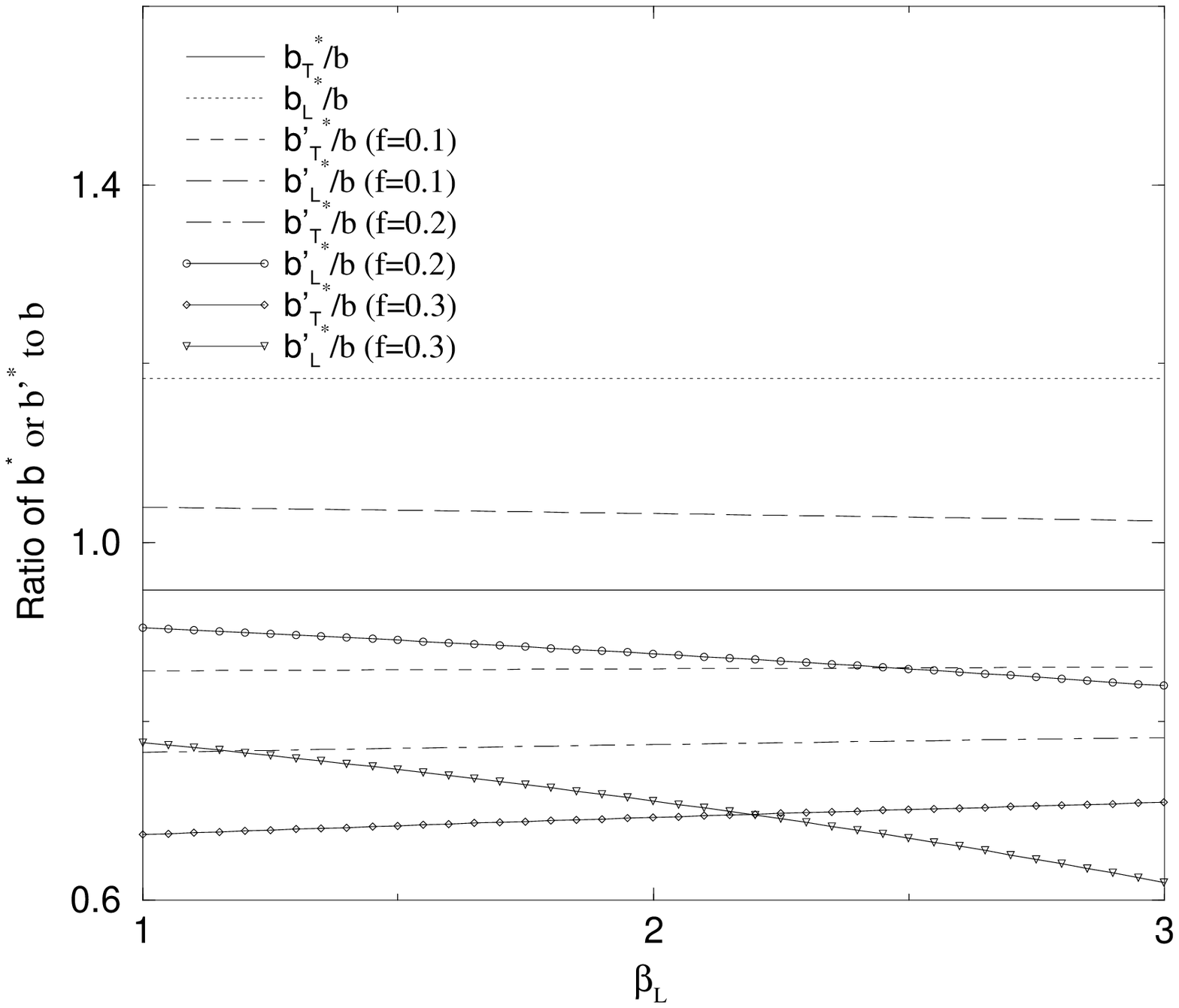,width=\linewidth}}
\centerline{Fig.2/Huang and Yu}

\newpage
\centerline{\epsfig{file=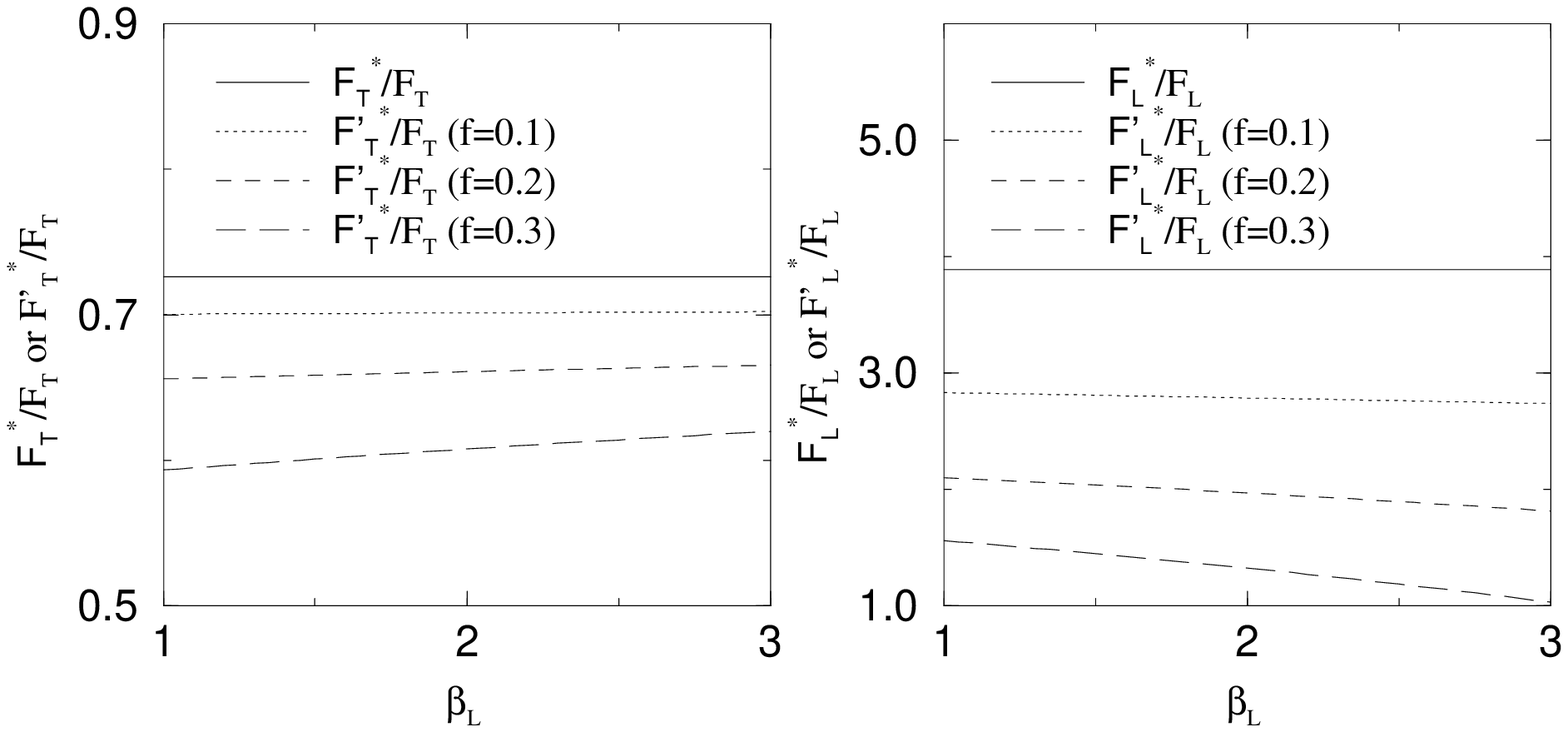,width=\linewidth}}
\centerline{Fig.3/Huang and Yu}

\end{document}